\documentstyle[pra,aps,twocolumn,graphics,epsfig]{revtex}
\def\kK{\mathcal{K} \displaystyle}
\def\kP{\mathcal{P} \displaystyle}
\begin{document}
\title{``Interaction-Free'' Imaging}
\author{Andrew G. White, Jay R. Mitchell, Olaf Nairz$^{*}$, and Paul G.  Kwiat}
\address{Los Alamos National Laboratory, P-23, MS-H803, Los Alamos,
NM 87545, USA\\
Fax: +1 (505) 665 4121, Phone: +1 (505) 665 3365,
e-mail: andrew.white@lanl.gov\\
$^{*}$ Institut f\"{u}r Experimentalphysik, Universit\"{a}t Innsbruck,
Technikerstrasse 25, 6020 Innsbruck, Austria}
\date{\small submitted 18th December 1997, accepted 9th March 1998}
\maketitle

\begin{abstract}
Using the complementary wave- and particle-like natures of photons, it
is possible to make ``interaction-free'' measurements where the
presence of an object can be determined with no photons being
absorbed.  We investigated several ``interaction-free'' {\it
imaging\/} systems, i.e.  systems that allow optical imaging of
photosensitive objects with less than the classically expected amount
of light being absorbed or scattered by the object.  With the most
promising system, we obtained high-resolution (10~$\mu$m),
one-dimensional profiles of a variety of objects (human hair, glass
and metal wires, cloth fibers), by raster scanning each object through
the system.  We discuss possible applications and the present and
future limits for interaction-free imaging.\\
\\
PACS number(s): 03.65.-w, 03.65.Bz, 42.50.-p, 42.25.Hz

\end{abstract}

\section{Introduction}
For most of us, our intuition of how the world works is grounded in
everyday experience, and so is necessarily classical.  Since its
earliest days, the field of quantum mechanics has been characterized
by predictions and apparent paradoxes that run counter to our natural
intuition.  However in remarkably short order, the practitioners of
quantum mechanics developed new intuitions \cite{CopenhagenMWetc}.
One of the widely accepted tenets of this new intuition is that in
quantum mechanics every measurement of a system disturbs the state of
that system (unless the system is already in an eigenstate of
the measurement observable).

Yet over the years a number of works have tested this new
intuition.  In 1960 Renninger showed that the state of a quantum
system could be determined via the {\it nonobservance\/} of a
particular result, i.e., the absence of a measurement or observation
can lead to definite knowledge of the state of the system
\cite{Renninger}.  In 1981 Dicke considered {\it ``interaction-free
quantum measurements''\/} where energy and/or momentum is transferred
from a photon to a quantum particle by the {\it nonscattering\/} of
the photon by the particle \cite{Dicke}.  And in 1993 Elitzur \&
Vaidman \cite{EV} showed that an arbitrary object (classical or
quantum) can affect the interference of a single quantum particle with
itself - the {\it noninterference} of the particle allows the presence
of the object to be inferred without the particle and object ever
directly ``interacting'' \cite{omit}.  In the Elitzur-Vaidman (EV)
interaction-free measurement (IFM) scheme, the measurement is
interaction-free at most half of the time.  Such experiments
were first performed in 1994 by Kwiat et  al.
\cite{IFMprl}, and later repeated as part of a public demonstration in
the Netherlands \cite{IFMrgy}.  Refs.~\cite{IFMprl,IFMmotirr} also 
proposed several schemes for high-efficiency IFM's: the fraction 
of IFM's exceeds one half, and in principle can be made arbitrarily 
close to unity, i.e.  the probability of absorption can be made arbitrarily 
close to zero.  In the first experiment using a high-efficiency 
system, Kwiat et al.  demonstrated the feasibility of performing IFM's 
up to 85\% of the time \cite{IFMhieta}.  The possibility of detecting 
the presence of an object without {\it ever\/} interacting with it led 
to the suggestion of {\it interaction-free imaging\/} (IFI) 
\cite{IFMsciam}, e.g., optical imaging of photosensitive objects with 
much less than the classically expected amount of light being absorbed 
or scattered by the object.  As one of the current limitations to 
imaging biological systems is power-induced optical damage the 
possibility of evading this limitation via interaction-free imaging 
bears further investigation.

While we realize that the best advantage of IFM techniques is 
realized in high-efficiency schemes, for the sake of conceptual and 
experimental simplicity, we consider in this paper only devices based 
on the EV scheme, that is, intrinsically low-efficiency devices.  
Specifically, we describe investigations of several possible 
interaction-free imaging devices, present experimental results from 
the most promising of these, and explore present and future limits to 
practicable IFI devices.  With these preliminary devices we obtained 
one-dimensional profile images - the objects were raster scanned 
through the beam of an interaction-free measurement system.  To obtain 
high spatial resolution, the beam at the imaging point is focused to a 
small size.

\begin{figure}
\begin{center}
\epsfxsize=\columnwidth
\epsfbox{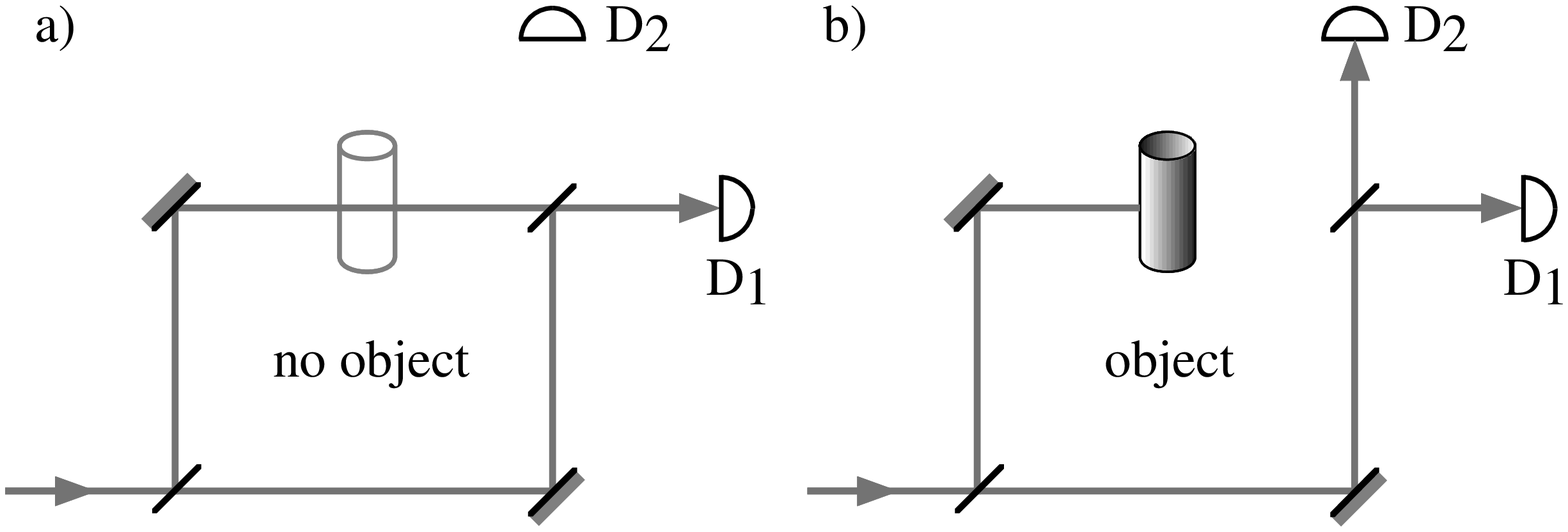}
\end{center}
\footnotesize
\caption{Elitzur-Vaidman scheme for interaction-free measurements. a) no 
object: the photon interferes with itself and no counts are detected 
at D$_{2}$.  b) object in one arm of the interferometer: the 
interference is destroyed and counts are detected at D$_{2}$ (counts 
detected one quarter of the time for 50/50 beamsplitters).}
\label{fig:EV}
\end{figure}
Figure \ref{fig:EV} shows the canonical EV scheme: a single photon
sent through a Mach-Zehnder interferometer.  The interferometer is set
so that, if no object is present, all of the light is output to port
1, none to port 2.  (Complete destructive interference is always
possible if the transmittance (reflectance) of the recombining
beamsplitter equals the reflectance (transmittance) of the first beam
splitter.) The probability of photon counts at detector 1 is thus
unity, whilst that at detector 2 is zero, i.e.  $\rm{P}(\rm{D}_{1})=1$
\& $\rm{P}(\rm{D}_{2})=0$.  If an opaque object is placed in one arm
of the interferometer the interference is destroyed.  The probability
of the photon being reflected by the first beamsplitter and thus
directed onto and being absorbed by the object is
$\rm{P}_{abs}=\rm{R_{1}}$.  The probability of detection at detector
1, i.e.,  the photon being transmitted by the first beamsplitter and
reflected by the second beamsplitter, is
$\rm{P}(\rm{D}_{1})=\rm{T_{1}R_{2}}$.  Note that in this {\it
no-result\/} case we gain no information on the presence of the object
- this detector can fire whether the object is there or not.  (The
high-efficiency schemes do not suffer this ambiguity
\cite{IFMprl,IFMhieta}).  The probability of detection at detector 2,
i.e.,  the photon being transmitted through both beamsplitters, is
$\rm{P}(\rm{D}_{2}) \equiv \rm{P}_{ifm}=\rm{T_{1}T_{2}}$ (we label
detector 2 the IFM detector: $\rm{D}_{2} \equiv \rm{D}_{ifm}$).
On the occasions that detector 2 fires we know that there is an object
in the interferometer arm, and we know that no photon was absorbed
since we only sent a single photon into the interferometer.  The
presence of the object has been determined without direct interaction
between the detected photon and the object.

The ``efficiency'' of an IFM device, that is, how often the device is
likely to make an interaction-free as opposed to an interaction-full
measurement, is defined as \cite{IFMprl}:
\begin{equation}
\eta = \frac{\rm{P}_{ifm}}{\rm{P}_{ifm}+\rm{P}_{abs}}.
\label{eqn:eta1}
\end{equation}
Assuming lossless beamsplitters, in the EV system considered here
this becomes
\begin{equation}
\eta = \frac{\rm{T_{1}T_{2}}}{\rm{T_{1}T_{2}}+\rm{R_{1}}}.
\label{eqn:eta2}
\end{equation}
If we add the condition that the transmittance of the second
beamsplitter is $\rm{T_{2}}=\rm{R_{1}}$ then
\begin{equation}
\eta = \frac{\rm{T_{1}}}{1+\rm{T_{1}}} = \frac{1-\rm{R_{1}}}{2-\rm{R_{1}}},
\label{eqn:eta3}
\end{equation}
and we see that $\eta \rightarrow 0.5$ as $R_{1} \rightarrow 0$.  Note
that {\it no-result\/} measurements (from detector 1) are not
considered, as we do not mind if a photon propagates through the
system and is neither absorbed by the object nor detected at detector
2.  For a balanced interferometer, where $R_{1}=T_{2}=0.5$ and
the intensities in both arms are equal, the probability of an
interaction-free measurement is at a maximum, $\rm{P}_{ifm}=0.25$;
however, the efficiency is only $\eta=0.33$ -- as the efficiency
increases there are more no-result measurements and the probability of
an IFM measurement actually decreases.  We stress that regardless of
the efficiency, when a single photon is detected at detector 2, that
particular measurement {\it is completely interaction-free\/}, as the
object has been detected yet the photon was not absorbed by the
object.  The efficiency only relays the {\it ratio\/} of
interaction-free to interaction-full and interaction-free
measurements: each {\it individual\/} single photon measurement is
either no-result, interaction-free, or interaction-full.

Single-photon experiments are more demanding than typical continuous
wave (cw) experiments in that they require special detectors, very low
background light levels, and so on.  Fortunately, it is not necessary
to use single photons to analyze and compare various interaction-free
imaging schemes.  The probability, $\rm{P}_{\rm{event}}$, of a
detection event in the single photon regime is related to the relative
intensity of that event in the cw regime:
\begin{equation}
\rm{P}_{\rm{event}} = \frac{\kP_{\rm{event}}}{\kP_{\rm{0}}},
\label{eqn:Prob vs Power}
\end{equation}
where $\kP_{\rm{0}}$ is the cw power incident to the interferometer
and $\kP_{\rm{event}}$ is the cw power detected at the event port
(i.e.  ports 1 or 2, or absorbed by the object).  All the experiments
presented in this work were done in the cw regime.  Obviously in this
regime no measurement is interaction-free: with many photons
simultaneously incident on the interferometer some can be absorbed by
the object whilst others can exit via port 2.  However, according to
the standard rules of quantum mechanics, by measuring the relative
intensity of light at a given port (as described in Eqn.~\ref{eqn:Prob
vs Power}) we can calculate the probability of an event at that port
in the single-photon regime.  In other words our evaluations in the cw
regime would be identical if performed with a single photon source and
detectors.

\section{Experiments}

\subsection{Imaging systems}
\label{imaging systems}
\begin{figure}
\begin{center}
\epsfxsize=\columnwidth
\epsfbox{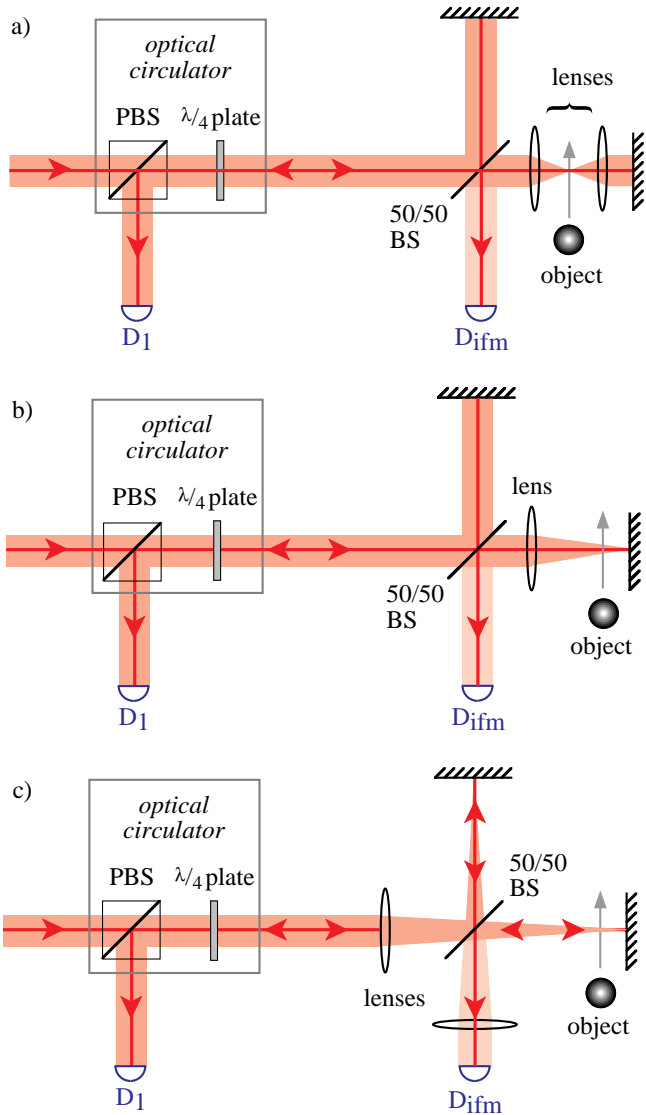}
\end{center}
\footnotesize
\caption{Conceptual layout of three interferometer configurations.  a) 
Michelson with two lenses in imaging arm, focus in free space.  b) 
Michelson with one lens in imaging arm, focus at end mirror.  c) 
Michelson with no lenses in interferometer.  Beamsplitters are all 
$\mathrm{R}=0.5$, so the maximum possible interaction-free efficiency in 
these configurations is $\eta=0.33$.}
\label{fig:michelson}
\end{figure}

\begin{figure}
\begin{center}
\epsfxsize=\columnwidth
\epsfbox{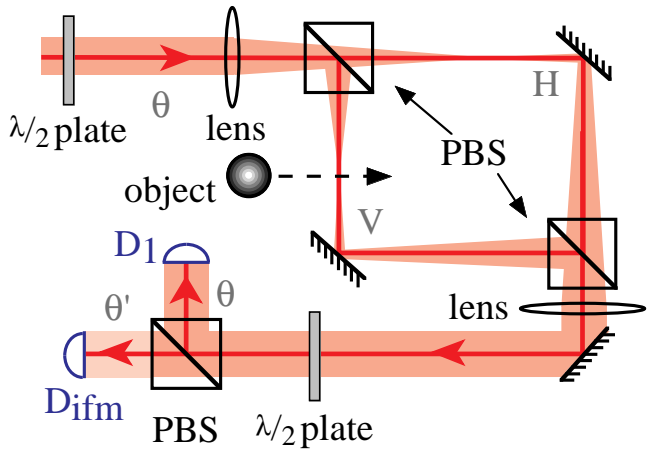}
\end{center}
\footnotesize
\caption{Polarizing Mach-Zehnder. PBS $\equiv$ polarizing 
beamsplitter.  $\lambda/2$ $\equiv$ half-wave plate at 670 nm.  
The locking laser (not shown) entered from the top port of the first 
PBS and exited from the side port of the second.}
\label{fig:MZ}
\end{figure}

Interaction-free imaging requires an instrument with high contrast
interference, in order to give low-noise interaction-free
measurements, and an accessible and small beam waist, to allow fine
resolution raster scanning of an object.  In all, four imaging systems
were investigated experimentally.  The first three systems were
variations on a Michelson interferometer (Fig.~\ref{fig:michelson}),
the last, a Mach-Zehnder interferometer (Fig.~\ref{fig:MZ}).  For all
systems the imaging beam was the output of a diode laser (1~mW @
670~nm, Thor Labs, Model 0220-999-0, circular output beam) that was
expanded and collimated by a telescope and then apertured with an
iris.  The detector was a calibrated photodetector (Newport 818-UV,
used with 1835-C power meter).

The first imaging system was a Michelson with two lenses (5x
microscope objectives), Fig.~\ref{fig:michelson}a.  This design had an
accessible beam waist between the two lenses.
Unfortunately, in practice it had very poor fringe visibility, as the
system was very sensitive to alignment mismatch between the lenses
(due to coma, astigmatism, etc.).  Given the poor performance, no data
was taken with this system.

The second system was a Michelson with a single lens in the imaging
arm, focused so that the waist was at the end mirror,
Fig.~\ref{fig:michelson}b.  As the beam was spatially inverted in the
imaging arm, but not the other, it was still difficult to get high
fringe visibility since we did not have the necessary, highly spatially
symmetric, wavefront.  Further, the waist was no longer
easily accessible: due to mechanical constraints, in practice it
was only possible to get an object to within $\sim 200 \
\mu$m of the waist.  Again, no data was taken with this system.

In the third system the lens was removed from within the
interferometer and placed before the first
beamsplitter, Fig.~\ref{fig:michelson}c.  This was the best of the
three Michelson systems that we considered, in that it had good fringe
visibility (in excess of 90\%) because both beams undergo the same spatial
inversion at their respective mirrors.  However, as for system 2, it
is not possible to image exactly at the waist.

The Michelson systems were investigated chiefly because of the
perceived advantages of their relative ease of alignment.  However,
regardless of the exact configuration, they all have one feature that
complicates interpretation of imaging data: the beam passes through
the object {\it twice\/}.  If the object is semi-transparent, then
twice the actual loss is experienced, and still further analysis of an
image is required.  Furthermore, a subtle effect means that any data
from systems 2 or 3 must be very carefully interpreted.  Consider the
following argument.  In system 3 (Fig.~2c) let half the beam be
blocked in the imaging arm at a point just after the beamsplitter.
The remaining half of the beam is focused onto the end mirror and
returns on the {\it other\/} side of the beam, where it too is
absorbed by the initial block.  Thus by blocking only half the beam in
the imaging arm, all the light in that arm is absorbed (neglecting
diffraction), and the interference is totally destroyed.  This effect
does not occur if the beam is blocked at the waist.  However, as
mentioned above, we could not image precisely at the waist.  In system
3 we typically imaged at around one Rayleigh range, i.e.,  in a region
somewhere between the farfield and the waist, meaning that this
half-beam effect is occurring to some degree.  The interpretation of
the data is then nontrivial: a full calculation accounting for the double
Fresnel edge-diffraction would be necessary.

The fourth imaging system was a Mach-Zehnder configuration, used to
obtain all the data presented here.  With this system it is easy to
arrange for an accessible beam waist in free space, and the beam only
passes through the object once.  Further, it was experimentally
necessary to lock the interferometers so that one port, the IFM port,
was at a null.  This was done with an additional laser (a HeNe at 632
nm) and a simple fringe slope locking system.  Incorporation of the
locking laser into a Michelson configuration was difficult due to the
intrinsic space constraints of that design; incorporation into a
Mach-Zehnder configuration was trivial - the empty ports of the
interferometer were utilized.

Fig.~\ref{fig:MZ} shows the Mach-Zehnder configuration: a {\it
polarizing\/} interferometer, which allows effective tuning of the
beamsplitter reflectances.  This configuration operates as follows:
the first half-wave plate ($\lambda/2$) is set so that the light input
to the interferometer is linearly polarized at $\theta$ from the
vertical axis.  The first polarizing beamsplitter (PBS) splits the
light into its horizontal ($\mathrm{T_{1}} = \mathrm{sin}^{2} \theta$)
and vertical ($\mathrm{R_{1}} = \mathrm{cos}^{2} \theta$) components
(for example, $\theta=45^{\circ}$ gives $\mathrm{R_{1}}=0.5$).  If no object
is present, the second PBS recombines the beams to the original
$\theta$ polarization, which is then is rotated back to the vertical
by the second $\lambda/2$ plate, so that the light is always detected
at $\mathrm{D}_{1}$.  If an object is present, however, the
interference is modified or destroyed.  In the latter case, only the
horizontal component is transmitted by the interferometer, the
vertical component being absorbed by the object.  (In quantum terms,
only the probability amplitude of the horizontal polarization path
contributes to the final probabilities).  The horizontally polarized
output is rotated towards the vertical axis by the second $\lambda/2$
plate, so that some counts occur at $\mathrm{D}_{ifm}$
($\mathrm{T_{2}} = \mathrm{cos}^{2}
\theta$) - these counts are the interaction-free measurements.  As with
the most successful Michelson system, the focusing lens
($\mathrm{f}=60$mm) was outside the interference region.

\subsection{Imaging results}
\begin{figure}
\begin{center}
\epsfxsize=\columnwidth
\epsfbox{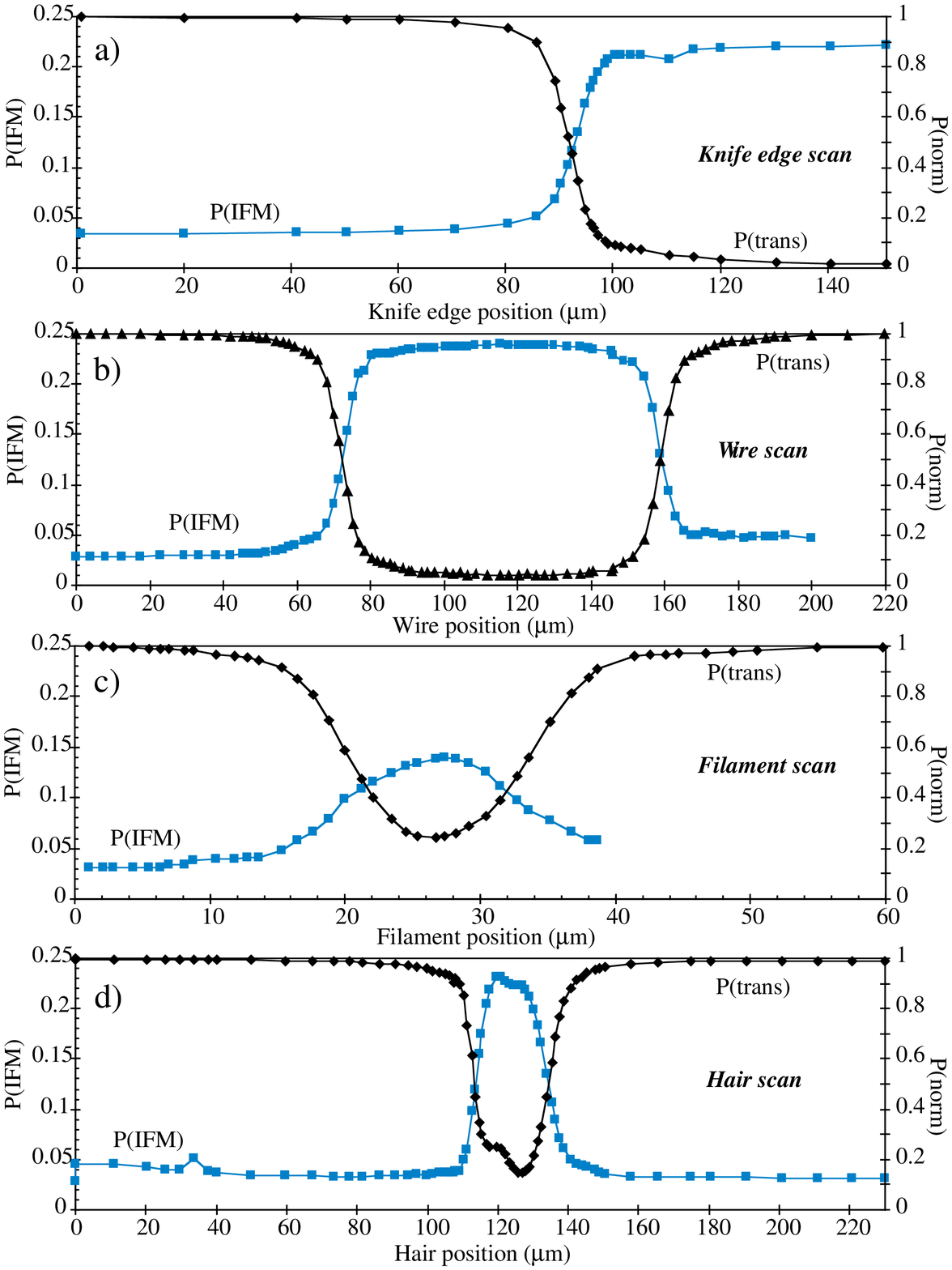}
\epsfxsize=\columnwidth
\epsfbox{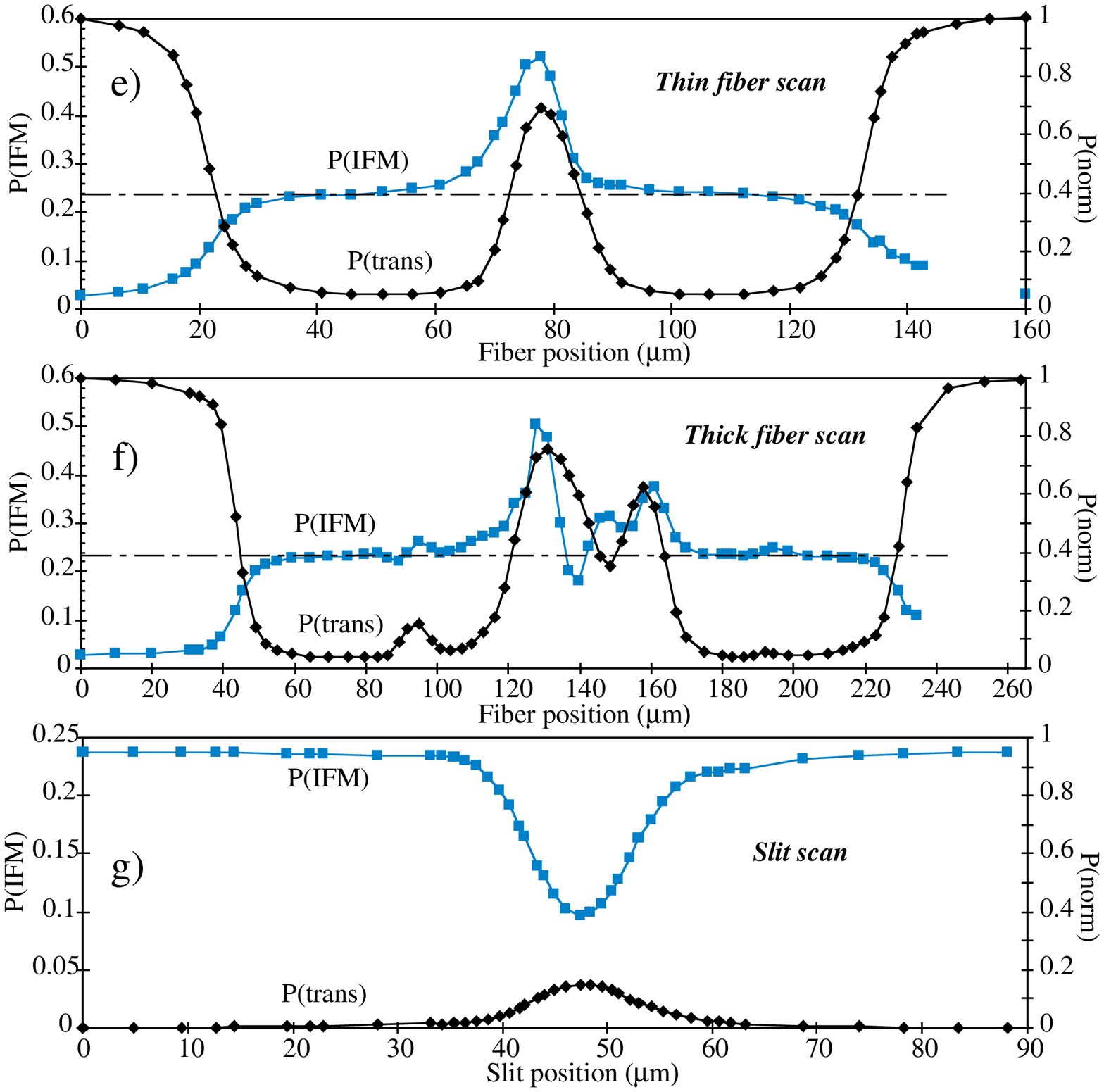}
\end{center}
\caption{Transmission and interaction-free images of various objects.
a) knife edge b) metal wire c) cloth filament d) human hair e) thin 
optical fiber f) thick optical fiber g) slit (absence of object). 
Note variation in scale on position axes.} 
\label{fig:scans}
\end{figure}

We performed one-dimensional scans of a variety of different objects,
including a simple knife-edge, human hair, metal wire, cloth and
optical fibers, and a narrow slit (the absence of an object).
Typical results are shown in Fig.~\ref{fig:scans}, these being
obtained for $\mathrm{R} \simeq 0.5$, i.e.,  input light polarized at
$\simeq 45^{\circ}$ and analyzing at $\simeq -45^{\circ}$.

The objects were scanned stepwise through the beam using a motorized
translation stage incorporating a high-resolution ($0.0555 \ \mu$m)
encoder.  At each step two measurements were recorded: the first was
an interaction-free measurement --- monitoring the dark port of the
interferometer (analyzing at $-45^{\circ}$) for an inhibition of the
interference; the second measurement was a normalized transmission
scan obtained by blocking the interferometer arm that did
not contain the object and measuring at the no-result port (detector
$\rm{D}_{1}$).  These are respectively the lefthand ($\rm{P}_{ifm}$)
and righthand ($\rm{P}_{norm}$) ordinates of Fig.~\ref{fig:scans}.
Note that $\rm{P}_{norm}$ is the probability of a photon being
transmitted through the object {\it outside\/} of the imaging system,
i.e.,  just the normal transmittance curve for the object.  The
probability that a photon is absorbed by the object when it is in the
imaging system is given by $\rm{P}_{abs}$, where
\begin{equation}
\rm{P}_{abs} = R_{1} \left[ 1-\rm{P}_{norm} \right].
\label{eqn:Pabs vs Pnorm}
\end{equation}

The knife-edge profile, Fig.~\ref{fig:scans}a, is used to measure the
resolution of the system.  Since the knife edge certainly has a
step-function profile on the micron scale, the rounding of the edges
on the scans is necessarily due to the spot size of the beam.  Taking
the derivative to obtain a Gaussian-like function, we infer a FWHM
spot size of $9.1 \pm 0.3 \ \mu$m.  The Rayleigh resolution of the
system is thus given by $10.7 \pm 0.3 \ \mu$m \cite{MellesGriot}.  The
difference between this and the theoretical value of
$\mathrm{d}=9.8$~$\mu$m (see Appendix 1) is probably due to the
non-ideal beam-quality --- despite aperturing down, the beam was still
not entirely spatially uniform.

The path of the knife edge through the beam is shown by the
transmission scan: the beam was initially unblocked ($x<60 \ \mu$m)
and the knife edge was scanned through until the beam was totally
blocked ($x>130 \ \mu$m).  In principle $\rm{P}_{ifm}=0$ in the
absence of the knife edge; however, in practice it is not, as shown by
the value $\rm{P}_{ifm}=0.035$ in Fig.~\ref{fig:scans}a.  This
background noise is from light leaking through the ``dark'' port due
to the imperfect fringe visibility ($\rm{V}=0.933$ for this scan),
and can be thought of as the ``dark noise'', $\sigma$, of the
interaction-free detector (for this scan
$\sigma=(1-\rm{V})/(1+\rm{V})=3.5$\%).  For the remainder of the
scans, the visibility was improved to reduce the noise, which varied
between $2.0-3.2$\%.

In the simple Mach-Zehnder EV scheme described in the introduction,
the IFM probability is set by the transmittance of the two
beamsplitters in the interferometer; in the polarizing Mach-Zehnder
this is instead the transmittance of the first polarizing beamsplitter
and the transmittance of the analyzing beamsplitter after the second
half-wave plate.  The exact values of these transmittances for a given
experiment can be inferred from the ratios of measurements at
$\rm{D}_{1}$ \& $\rm{D}_{ifm}$ for both the transmission and the IFM
scans when the object is fully blocking the beam.  Recall that when an
object is fully blocking the beam the expected IFM probability is the
product of these transmittances.  For the knife edge scans
$\rm{T}_{1}=0.467$, $\rm{T}_{2}=0.422$ and so
$\rm{P}_{ifm}^{th}=0.23$, in good agreement with the observed value of
$\rm{P}_{ifm}=0.22 \pm 0.01$ on the right hand side of the IFM scan in
Fig.~\ref{fig:scans}a (the error on each data point in the IFM scans
is typically $\pm 4$\% of the value of that point).

Fig.~\ref{fig:scans}b is a profile of a metal wire.  The diameter
(FWHM) of the wire was estimated from both the transmission ($96.6\pm
1.0 \ \mu$m) and IFM ($96.6\pm 1.0 \ \mu$m) scans and was in good
agreement with the width measured via a microscope ($95.5\pm 1.6 \
\mu$m) and diffraction of a laser beam ($97.0\pm 0.5 \ \mu$m).  A
larger wire was also scanned (not shown) and again agreement between
the transmission ($162.7 \pm 1.6 \ \mu$m), IFM ($160.2 \pm 1.6 \
\mu$m) scans and microscope ($159.1\pm 2.3 \ \mu$m) and diffraction
measurements ($159.5\pm 2.0 \ \mu$m) was very good.  This gives us
confidence that the system can be used to accurately profile opaque objects
at this scale.

For these scans the transmittances were adjusted ($\rm{T}_{1}=0.525$,
$\rm{T}_{2}=0.462$) to give a higher expected IFM probability,
$\rm{P}_{ifm}^{th}=0.24$.  Again this agrees with the actual IFM
values observed in the center of the IFM scan where the wire totally
obscures the beam.  The efficiency of the measurement in the central
region can be calculated directly from Eqn.~\ref{eqn:eta2}, we obtain
$\eta=0.34$.  Alternatively, $\eta$ can be calculated via
Eqn.~\ref{eqn:eta1}; however, this requires the probability of
absorption, $\rm{P}_{abs}$, which was not measured directly.
Fortunately, $\rm{P}_{abs}$ can be calculated from the measured value
of the normalized probability of transmission (see Eqn.~\ref{eqn:Pabs
vs Pnorm}).  In the central region $\rm{P}_{abs}=0.46$, again giving
an experimental efficiency of $\eta=0.34$.  The agreement between the
efficiency calculated only from the reflectances and the efficiency
calculated using the inferred absorption probability gives us
confidence in the experimental analysis.

Note that the noise on the IFM scan rises slightly towards the right
hand side of the scan, and that the IFM scan terminates before the
transmission scan.  This behavior is due to the non-ideal lock of the
interferometer: the system gradually drifted away from the dark
fringe, increasing the light and thus the noise through the dark port,
before finally losing lock, ending the IFM scan.  This behavior is
seen on several of the scans (b,c,e,f) and highlights the importance
of a robust locking scheme in future IFI systems.

Fig.~\ref{fig:scans}c is a profile of a cloth fiber.  The FWHM
diameter was measured to be $12.6 \pm 0.6 \ \mu$m (microscope) \&
$15.4 \pm 1.2 \ \mu$m (diffraction), respectively.  The difference
between these two measurements suggest that the fiber had a
non-uniform or non-isotropic (e.g.  elliptical) cross-section, and
that different sections or orientations of the fiber were measured by
the two different techniques, giving slightly different widths.  This
is further borne out by the FWHM diameters measured from the
transmission and IFM scans (16.3 $\mu$m \& 16.6 $\mu$m, respectively):
they are consistent with one another, are within one sigma of the
diffraction measurement, and differ significantly from the microscope
measurement.

A more important feature of this scan is that as the transmission
never drops to zero (i.e.  the cloth fiber is not fully opaque), the
probability of an interaction-free measurement never attains its
maximum value of one quarter.  At the minimum of transmission,
$\rm{P}_{norm}=0.24$, from which we expect, $\rm{P}_{ifm}^{th}=0.07$
(see Appendix 2 for calculating $\rm{P}_{ifm}$ from $\rm{P}_{norm}$).
Actually, the observed value was higher than this, $\rm{P}_{ifm}=0.14$.
A similar discrepancy is seen in the profile of a human hair,
Fig.~\ref{fig:scans}d.  Note the internal structure of the traces.  As
can be seen from the transmission scan, the hair is also not totally opaque
(the transmission never falls to zero, c.f.  scans a \& b), and
furthermore near the center of the hair ($x \simeq 119 \ \mu$m) more
light is transmitted than at the edges, particularly the right edge
($x=123 \ \mu$m).  Left of center, where the object is less
opaque and there is seemingly less chance of an IFM, one might
expect the IFM scan to drop accordingly; however, it clearly {\it
increases\/}.  This is even more striking in the profile of a thin
optical fiber, as shown in Fig.~\ref{fig:scans}e.  Here for two thirds
of the width of the fiber, the fiber is essentially opaque (due to
scattering and reflection from the curved surface of the fiber), and
$\rm{P}_{ifm}$ is near the expected value of $\rm{P}_{ifm}^{th}=0.23$.
However, in the middle of the fiber the transparency increases
notably, and $\rm{P}_{ifm}$ attains values of up to 0.52, exceeding
even the na\"{i}ve in-principle limit of 0.25.

In all three cases (scans c-e) we believe the increase in 
$\rm{P}_{ifm}$ is caused by the light transmitted through the object 
acquiring a relative phase shift, which changes the interference 
conditions and so causes the IFM port to no longer be at a dark 
fringe.  This is clearly an important phenomena in IFM measurements 
(and in fact, is present to an even greater degree in high-efficiency 
schemes \cite{IFMhieta}).  Consider, for example, imaging a completely 
transparent object ($\rm{P}_{norm}=1$; $\rm{P}_{abs}=0$) that 
introduces a $\pi$-phase shift: all the light is detected at the 
``dark'' port detector yielding a 100\% efficiency, i.e.  
$\rm{P}_{ifm}=1$ and $\eta=1$.  As the transparency of such an object 
is reduced, then $\rm{P}_{ifm}$ and $\eta$ decrease accordingly.  In 
the limit where the object is totally opaque we recover our familiar 
results of $\rm{P}_{abs}=R_{1}$, $\rm{P}_{ifm}=T_{1}T_{2}$ and $\eta$ 
as given by Eqn.~\ref{eqn:eta3}.  As soon as there is some probability 
that a photon can be transmitted through the object, it is no longer 
sensible to describe the measurements as interaction-free, and 
concepts and equations based on the assumption of detecting a wholly 
opaque object need to be used with care (see Section~\ref{discuss}).

But what exactly causes the phase shift?  As shown by the asymmetry of
the IFM scan in Fig.~\ref{fig:scans}e it is clearly associated with,
but not directly proportional to, the increase in transparency.  There
are several possible causes for both the phase and transparency
shifts: scattering and reflection from the object; the phase shift due
to passage through the object ($\phi=(2 \pi (n-1) D)/\lambda$, where
$D$ is the width of the object that the light passes through); and the
geometrical phase shift due to the additional focusing from a
semi-transparent cylinder (i.e.  the Guoy phase shift associated with
focused beams; approximately $\pi$ radians \cite{Siegman}).  In
Fig.~\ref{fig:scans}e $\rm{P}_{norm}=0.69$, from which we expect
$\rm{P}_{ifm}^{th}=0.007$, if there were no phase shift (see Appendix
2).  As the experimental value is $\rm{P}_{ifm}=0.52$, we calculate,
using Eqn.~\ref{eqn:appx2e} from Appendix 2, that the relative phase
shift for light passing through the center of the fiber is
104$^{\circ}$.

It is tempting to interpret the transmission scan of
Fig.~\ref{fig:scans}e as a straightforward image of the well-known
internal structure of an optical fiber, i.e.,  a core cylinder of glass
surrounded by a cladding cylinder of higher refractive glass.
However, given the opportunity for refraction and beam-steering,
things are not likely to be so straightforward, as borne out by the
profile of a thicker optical fiber in Fig.~\ref{fig:scans}f.  Here
there are 4 peaks in the transmission scan (two central, and two small
side peaks) and four corresponding features in the IFM scan.  These
are most likely due to guided and scattered light paths, and certainly
don't represent a simple profile of the core structure.  The shape of
the features in the IFM scan reflect that the phase shift across the
transmission peaks is large and non-uniform.

Finally, Fig.~\ref{fig:scans}f is the profile of the {\it absence\/} 
of an object, i.e., a slit.  The slit was constructed by aligning two 
razor-blade edges in close proximity.  Due to mechanical constraints 
the blade edges were not exactly parallel and the slit was marginally 
V-shaped.  From the transmission and IFM scans, we respectively infer 
a slit width of $13.1 \ \mu$m \& $12.5 \ \mu$m; from the diffraction 
measurement, a width of $19.2\pm 1.2 \ \mu$m.  It is probable that the 
difference was due to a slightly different vertical alignment of the 
slit with respect to the beam.  Note that this, combined with a small 
longitudinal shift from the waist position, may also explain the 
surprisingly low transmission (the slit was effectively nearly 
opaque).  The IFM scan is sensitive to small changes in the effective 
transparency of the object: when the object fully blocks the beam, 
$\rm{P}_{norm}=0$ ($\rm{P}_{abs}=0.49$) and $\rm{P}_{ifm}$ is at its 
maximum value, $\rm{P}_{ifm}=0.24$; a small change in the 
transparency, to $\rm{P}_{norm}=0.15$, $\rm{P}_{abs}=0.43$, leads to a 
much larger change in the IFM scan, $\rm{P}_{ifm}<0.097$ (which agrees 
within error with the expected value, $\rm{P}_{ifm}^{th}=0.094 \pm 
0.007$).  This sensitivity to small changes in transparency holds 
promise for high-relief interaction-free imaging of low-relief 
absorption objects, with much less than the classically necessary 
light flux.  The effect is in fact more pronounced in 
high-efficiency schemes, wherein it is possible for a given absorptive 
object to have a {\it lower\/} probability of absorption than another 
object with {\it lesser\/} intrinsic absorptance \cite{IFMhieta}.

As a final note, we point out that it may be possible to use the
current device to obtain information on the polarization properties of
objects.  As currently used, the object is interaction-free imaged by
purely vertically-polarized light (i.e.  {\it s}-polarized with
respect to vertically aligned objects); equally validly,
interaction-free imaging could be done in the complementary arm with
horizontally-polarized light (i.e.  {\it p}-polarized w.r.t vertically
aligned objects).  Fine polarization-dependent details could then be
brought out by looking at the difference of the two interaction-free
images.

\subsection{Approaching high efficiency}
In principle the measurements in the last section could have been made
at efficiencies higher than $\eta=0.33$ (up to $\eta=0.5$ in the EV
scheme).  However, there was a strong experimental reason that this
was not done.  As discussed previously, the probability of an IFM in
the EV scheme is actually highest when
$\mathrm{R}_{1}=\mathrm{T}_{2}=0.5$, i.e.  $\rm{P}_{ifm}=0.25$ and
$\eta=0.33$.  Because the IFM noise floor is a fixed value set by the
visibility of the interferometer ($\sigma \sim$2-3\% for our system) the
greatest
signal to noise ratio (and so the greatest detail) for IFM scans is attained
when $\rm{P}_{ifm}=0.25$.

\begin{figure}
\begin{center}
\epsfxsize=\columnwidth
\epsfbox{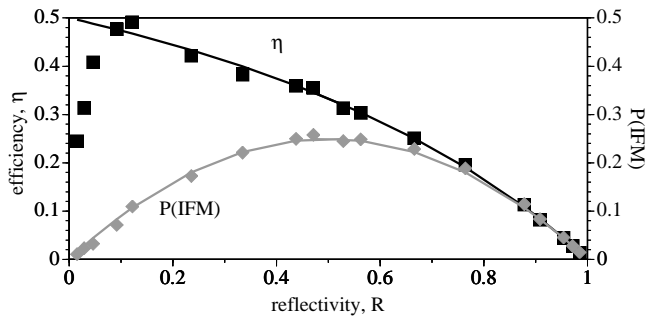}
\end{center}
\footnotesize
\caption{Efficiency and probability of an interaction-free measurement 
as a function of reflectance.  Diamonds and squares respectively 
represent experimentally measured values of $\rm{P}_{ifm}$ and 
efficiency $\eta$.  The unbroken lines are the theoretical curves.}
\label{fig:eff}
\end{figure}

To investigate this issue, $\rm{P}_{ifm}$ and efficiency were
measured for a range of reflectances.  An opaque object completely
blocked the imaging arm of the polarizing Mach-Zehnder. By
appropriately varying the angles of the two half-wave plates (see
Figure \ref{fig:MZ}) the reflectances were varied so that
$\mathrm{R}_{1}=\mathrm{T}_{2} \equiv \mathrm{R}$, where $0<\mathrm{R}<1$.
The results are shown in Fig.~\ref{fig:eff}.  For high reflectances,
$\rm{P}_{ifm} \simeq \eta$, as the probability of absorption is very
high.  $\rm{P}_{ifm}$ attains its maximum value at $\rm{R}=0.5$; in
the region $0.5<\mathrm{R}<1$, $\rm{P}_{ifm}$ decreases because the
probability of a no-result measurement increases.  (The $\rm{P}_{ifm}$
decrease is not reflected in the efficiency because by definition
$\eta$ depends only on the ratio of $\rm{P}_{ifm}$ to $\rm{P}_{ifm}$
\& $\rm{P}_{abs}$, and $\rm{P}_{abs}$ also decreases as $\mathrm{R}
\rightarrow 0$.)

The experimental values of $\rm{P}_{ifm}$ were calculated directly
from the output powers (as described by Eqn.~\ref{eqn:Prob vs Power}).
To obtain the experimental values for efficiency it was necessary to
know the values for $\rm{P}_{abs}$: these were calculated by assuming
that the sum of the absorption, interaction-free, and no-result powers
equaled the observed output power in the {\it absence\/} of an object.
The agreement between experiment and theory for $\rm{P}_{ifm}$ is
excellent

The agreement for the efficiency is also very good but breaks down
badly at low reflectances, when polarization cross-talk degrades the
efficiency.  Polarizing beam splitters (PBS) are designed to separate
an arbitrarily polarized beam into its horizontal and vertical
components.  The {\it cross-talk\/} of a PBS is the residual amount of
the orthogonal polarization on each ``pure'' output beam.  Thus at low
reflectances, where $\rm{P}_{abs}$ is in principle vanishingly small,
in practice it has a fixed value, set by the cross-talk.  The
undesirable consequence of this is that, while $\rm{P}_{ifm}$
decreases as $\rm{R} \rightarrow 0$, $\rm{P}_{abs}$ is fixed due to
the cross-talk, and thus the efficiency $\eta$ decreases sharply.  The
system only behaves as described by Eqns.~\ref{eqn:eta1}-\ref{eqn:eta3}
for reflectances above $\sim 10$\%.

\section{Discussion}
\label{discuss}
As has been touched upon in the experimental results section, 
semi-transparent objects necessarily force a reevaluation of what is 
meant by an ``interaction-free'' measurement.  The original central 
idea of interaction-free measurement was a totally opaque object 
causing the non-interference of a single photon \cite{EV}.  Classical 
objects can modify this effect if they are semi-transparent or 
diffract the light.  A transparent or semi-transparent object can 
phase-shift the light and modify the interference (we note in passing 
that such shifts can in principle yield information about the 
dispersive properties of the object).  However, even in the absence of 
such a phase shift, some interference will still occur, as any 
transmitted light may interfere with the light from the other arm of 
the interferometer.  Similarly, even a totally opaque object may allow 
interference if it diffracts light in such a way that it can overlap 
with light from the other arm.

Quantum objects may also be imaged by interaction-free detectors
\cite{IFMcat,IFMatom}.  For these objects, any forward scattering (be
it due to transparency, diffraction, re-emission, or some other
process) will allow some degree of interference.  Further, during
``interaction-free'' measurements of quantum objects, momentum and
energy can be transferred from the light to the object
\cite{Dicke,DanielPaulMe} if there is a forward scattering amplitude
from the object.  Energy and momentum transfer are unusual phenomena
indeed for an ``interaction-free'' measurement!

Accordingly, we reiterate that as soon as there is some probability
that a photon can be transmitted or diffracted by the object, it is no
longer sensible to describe the measurements as truly
interaction-free, at least in the original sense of the phrase.

An interaction-free measuring system can be thought of as a detector,
albeit an unusual one.  As with all detectors, IFM systems are
characterized in terms of their efficiency ($\eta$) and noise
($\sigma$).  The interaction-free imaging systems considered here are
extensions of this concept --- they are IFM detectors with fine spatial
resolution.  The ultimate limit to spatial resolution for any standard
optical detector is the diffraction limit: in the current system we
are still some way from achieving this limit ($\sim$ 10~$\mu$m vs
$\sim$ 0.6~$\mu$m).  In the future it may be better to avoid
polarization-based IFM detectors, since the polarization cross-talk
limits both the spatial resolution (see Appendix 1) and the minimum
noise of the detector.

For practical applications the greatest benefit will be obtained by
incorporating imaging into a high efficiency IFM
system ~\cite{IFMprl,IFMmotirr}, as it is only
in these systems that the chance of a photon interacting with the
object becomes vanishingly small.  Because such systems are in their
infancy, with current efficiencies of only $~60 - 85$\%
\cite{IFMhieta}, the issue of incorporating imaging into these systems
is non-trivial, and requires further research.  Aside from this,
further technical improvements are conceivable: for example, the
possibility of obtaining an image ``all at once'' using a large
diameter interrogating beam, and some sophisticated image-processing
algorithm to back-out from the interference pattern the image of the
object.

Classical objects that would benefit from the greatly reduced photon
flux of interaction-free imaging include: biological systems, such as
cells \cite{cell}, whose biological and chemical operation can change
as a function of light level; and cold atom clouds, which can
literally be blown apart from the photon flux of conventional imaging
systems.  Not only could a variety of ``delicate'' quantum objects
(such as trapped ions, Bose-Einstein condensates, or atoms in an atom
interferometer) be interaction-free imaged as well, but in high efficiency
systems the act of imaging can entangle the imaging photons and the
quantum object, creating novel quantum-mechanical states, such as
entangled ``Schr\"{o}dinger Cat'' states \cite{IFMcat}.

The difference between conventional and interaction-free measurements
of the {\it presence\/} of an object is that in the latter, in
principle the object can be detected with {\it no\/} photons
interacting with the object.  Similarly, the principal difference
between conventional and interaction-free {\it imaging\/} of an object
is the vastly reduced photon flux needed to obtain an image in the
latter.  Current photonic imaging systems (e.g., optical low coherence
reflectometry, OCLR) can have very high sensitivity (to opacity), say
one part in $10^{12}$ (-120 dB).  However, this is at the expense of
sending $10^{12}$ photons {\it through\/} the imaged object and having
at least one of those photons interact in a detectable fashion (e.g.
in OCLR, by backscattering) - of course the remaining photons can and
do interact with the object in a variety of ways (general scattering,
absorption, etc.).  In contrast to this, we suggest that a
high-efficiency interaction-free imaging system might attain high
sensitivity by having only a few photons interact with the object, the
rest remaining in the other arm of the interferometer - further
analysis is needed to quantify this.  In any event, it is clear that
the techniques of interaction-free measurements and imaging, presented
here and elsewhere, offer unique capabilities beyond those normally
considered in conventional optics.

\section{Acknowledgments}
We wish to acknowledge fruitful discussions with Anton Zeilinger,
Raymond Y.  Chiao, Morgan W.  Mitchell, Anders Karlsson and Gunnar
Bj\"{o}rk.  We thank Sky Frostenson for performing the diffraction
measurements of the objects.  O.N.'s participation was funded by the
Austrian Science Foundation (FWF project number S65-04).

\section{Appendix 1: Image resolution}
As discussed in Section~\ref{imaging systems}, it is desirable to have
a small beam waist in the region where the object is scanned, in order to
obtain high spatial resolution.  The diameter of a spot available from a
lens is given by:
\begin{equation}
\mathrm{d} = \kK \frac{\mathrm{f} \lambda}{\phi_{\mathrm{D}}},
\label{eqn:spot size}
\end{equation}
where $\mathrm{f}$ is the focal length of the lens, $\lambda$ is the
wavelength of the light, $\phi_{\mathrm{D}}$ is the diameter of the
clear aperture at the lens, and $\kK$ is a numerical factor that
depends on experimental conditions and whether the diameter under
consideration is the full-width half-maximum (FWHM) or the Gaussian
diameter (where the power has fallen to $1/\mathrm{e}^{2}$ of the
original value).  For a lens imaging an unapertured Gaussian beam, the
Gaussian diameter is given by $\kK=\mathrm{4}/\pi$.  However, the
output of our diode laser was not a clean Gaussian mode as it had
internal structure (e.g.  ``picket fencing'').  To reduce effects from
this structure, the beam was expanded to $\sim 25$~mm diameter
($1/\mathrm{e}^{2}$) and then a more spatially uniform subsection of
the beam was selected with an iris ($\phi_{\rm{D}}=5$mm) placed before
the imaging lens.  Under these conditions, the beam input to the iris
is approximately plane wave and the factor $\kK$ varies as a function
of the truncation of the initial beam, $\mathrm{T}$:
\begin{equation}
\mathrm{T} = \frac{\phi_{\mathrm{beam}}}{\phi_{\mathrm{iris}}}
\label{eqn:truncation}
\end{equation}
where $\phi_{\mathrm{beam}}$ is the $1/\mathrm{e}^{2}$ diameter of the
input beam and $\phi_{\mathrm{iris}}$ is the physical diameter of the
iris.  To calculate FWHM diameters, the factor $\kK$ is given by
\cite{MellesGriot}:
\begin{equation}
\kK = \mathrm{1.029} + \frac{0.7125}{(\mathrm{T}-0.2161)^{2.179}} -
\frac{0.6445}{(\mathrm{T}-0.2161)^{2.221}}.
\label{eqn:K factor}
\end{equation}
Taking the 5~mm iris diameter as the clear aperture of the lens, the
60~mm focal length lens and initial beam diameter of
$\phi_{\mathrm{beam}}=25$~mm yield a factor $\kK=\mathrm{1.03}$.  Thus
the predicted minimum spot size for the system is
$\mathrm{d}=8.3$~$\mu$m (FWHM), and the predicted minimum resolution
(as defined by the Rayleigh criterion \cite{MellesGriot}) is
$\mathrm{d_R}=9.8$~$\mu$m.

It is possible in principle to attain a smaller spot size by
increasing the diameter of the iris.  However,
in practice this was limited by two experimental factors: the
non-uniform beam and the angle-dependent cross-talk at the polarizing
beam splitters.  As mentioned above, the output beam from the diode
laser contained spatial structure.  If the beam was unapertured the
diffraction of this structure meant that the achievable fringe
visibility was quite low, $<60$\%; by aperturing a uniform subsection
of the beam the fringe visibility was improved to $~95$\%.  This
aperturing was merely for the sake of convenience, and could have been
avoided by a suitable mode cleaning system.  However, the second
effect could not have been so avoided.  The cross-talk on the
polarizing beam splitters is a minimum when the beam passing through
the device is collimated.  As the beam becomes strongly diverging or
converging (as was the case in our experiment), the amount of
cross-talk increases rapidly.  This behavior occurs both for interface
PBS's (such as the cube PBS's we used) and for bulk PBS's (e.g.
calcite prisms).  Our aperture size of 5 mm was thus chosen to give an
acceptable trade-off between imaging spot size and polarization
cross-talk.

\section{Appendix 2: Calculating $\rm{P}_{ifm}$ from $\rm{P}_{norm}$.}
>From the normalized transmission probability, $\rm{P}_{norm}$, it is
straightforward to calculate the expected interaction-free measurement
probability, $\rm{P}_{ifm}$, as long as the reflectances
($\rm{R}_{1}$, $\rm{R}_{2}$) of the interferometer are known.  Consider
inputting linearly polarized light (at $\theta_{1}$) to the polarizing
Mach-Zehnder described in section~\ref{imaging systems}.  We use a Jones
matrix description, where for example, light linearly polarized at an
angle $\theta$ with respect to the vertical axis is described as
\begin{equation}
\left[
\begin{array}{c}
\mathrm{sin} \theta \\
\mathrm{cos} \theta
\end{array}
\right].
\label{eqn:appx2a}
\end{equation}
After passing through the interferometer the light is described by
\begin{equation}
\left[
\begin{array}{cc}
t \mathrm{e}^{\mathit{i} \phi} & 0 \\
0 & 1
\end{array}
\right] . \left[
\begin{array}{c}
\mathrm{sin} \theta_{1} \\
\mathrm{cos} \theta_{1}
\end{array}
\right],
\label{eqn:appx2aa}
\end{equation}
where $t$ is the real part of the free-space
transmittivity of the object, and $\phi$ is the phase shift
that the light acquires in its passage through the object :
\begin{equation}
\mathrm{P}_{norm} = \mathit{t}^{\mathrm{2}}.
\label{eqn:appx2b}
\end{equation}
After passing through the analyzer at angle ($\theta_{2}$), the probability
of an
interaction-free measurement, $\rm{P}_{ifm}$, is
\begin{eqnarray}
\rm{P}_{ifm} & = & \left| [\mathrm{sin} \theta_{2}, -\mathrm{cos} \theta_{2}]
. \left[
\begin{array}{cc}
t \mathrm{e}^{\mathit{i} \phi} & 0 \\
0 & 1
\end{array}
\right]
. \left[
\begin{array}{c}
\mathrm{sin} \theta_{1} \\
\mathrm{cos} \theta_{1}
\end{array}
\right] \right| ^{2} \nonumber \\
& = & |t \mathrm{e}^{\mathit{i} \phi} \ \mathrm{sin} \theta_{1}
\mathrm{sin} \theta_{2} - \mathrm{cos}
\theta_{1} \mathrm{cos} \theta_{2}|^{2}.
\label{eqn:appx2c}
\end{eqnarray}
>From this, and remembering that the effective reflectance of the first
beamsplitter is $\rm{R}_{1} = \mathrm{cos}^{2} \theta_{1}$, that the
transmittance of the analyzer is $\rm{T}_{2} = \mathrm{sin}^{2}
\theta_{2}$, and that $\rm{R}_{i} + \rm{T}_{i} = 1$, we rewrite
Eqn.~\ref{eqn:appx2c}:
\begin{eqnarray}
\rm{P}_{ifm} & = & \rm{R}_{1} \rm{R}_{2} + \rm{T}_{1} \rm{T}_{2}
\rm{P}_{norm} \nonumber \\
             & \phantom{=} & - 2 \mathrm{cos} \phi \sqrt{\rm{R}_{1} \rm{R}_{2}
                             \rm{T}_{1} \rm{T}_{2} \rm{P}_{norm}} ,
\label{eqn:appx2d}
\end{eqnarray}
which relates $\rm{P}_{ifm}$ to $\rm{P}_{norm}$.  For the case of
50/50 beamsplitters (i.e., input light polarized at $\theta_{1} =
45^{\circ}$),
this reduces to:
\begin{equation}
\rm{P}_{ifm} = \frac{1 + \rm{P}_{norm} - 2 \mathrm{cos} \phi
\sqrt{\rm{P}_{norm}} }
{4} = \frac{|1-\mathit{t} \mathrm{e}^{\mathit{i} \phi}|^{2}}{4}
\label{eqn:appx2e}
\end{equation}
In this case, if the object is totally opaque, ($\rm{P}_{norm}=0$), 
then $\rm{P}_{ifm} = 1/4$, as expected.  (And of course, if the object 
is absent then $\rm{P}_{norm}=1$, $\phi=0$, and $\rm{P}_{ifm} = 0$.) 
Of course, in the EV scheme considered here, the probability of the 
object absorbing a photon is independent of the interference 
conditions, and in all cases is given by $\rm{P}_{abs}=\mathit{t}^{2} 
\rm{R_{1}}$.

\newpage
\footnotesize
\onecolumn
\begin{table}
\caption{Object widths: inferred from ``interaction-free'' and
normalized transmission scans, measured with microscope and
diffraction.  The uncertainty of the widths from the IFM and
transmission scans are approximately $\pm 1$\%, except for the cloth
filament where they are approximately $\pm 2$\%. \\}
\begin{tabular}{|l|c|c|c|c|} 
Object & Width inferred & Width inferred from & Width measured &
Width measured \\ 
 & from IFM scan & transmission scan &
 by microscope & via diffraction \\ 
               & [$\mu$m] & [$\mu$m] & [$\mu$m] & [$\mu$m]       \\ \hline
               &       &       &                &                \\ 
thin metal     & 95.3  &  96.6 &	$95.5\pm 1.6$ & $97.0\pm 0.5$  \\ 
wire           &       &       &                &                \\ 
thick metal    & 160.2 & 162.7 & $159.1\pm 2.3$ & $159.5\pm 2.0$ \\ 
wire           &       &       &                &                \\ 
cloth          & 16.6  &  16.3 & $12.6\pm 0.6$  & $15.4\pm 1.2$  \\ 
filament       &       &       &                &                \\ 
human hair     & 22.8  &  24.7 & $25.1\pm 0.9$  & $26.2\pm 0.6$  \\ 
               &       &       &                &                \\ 
thin optical   & 125.7 & 123.9 & $123.5\pm 1.9$ & $123.2\pm 3.6$ \\ 
fiber          &       &       &                &                \\ 
thick optical  & 208.0 & 207.5 & $207.9\pm 3.0$ & $208.3\pm 2.5$ \\ 
fiber          &       &       &                &                \\ 
slit           & 12.5  & 13.1  &       -        & $19.2\pm 1.2$  \\ 
               &       &       &                &                \\ 
%
\end{tabular}
\label{tab:photodiodes}
\end{table}

\begin{thebibliography}{123}
\bibitem{omit} There has been some controversy and
misunderstanding concerning what is meant by ``interaction'' in the
context of ``interaction-free'' measurements.  In particular, we
stress that there must be a coupling (interaction) term in any
Hamiltonian description of the IFM system.  For brevity's sake,
throughout the rest of this paper we omit the quotes from
``interaction-free'', although, as we shall see, they are often
justified.
\bibitem{cell} As can be seen from the scan of the thick optical
fiber, imaging thick objects that can steer the light is non-trivial.
Our current system is better suited to imaging thin objects where this
does not occur to any great extent, such as some varieties of cells.
\end{thebibliography}
\end{document}